\documentclass[apj]{emulateapj}

\usepackage{graphicx} 
\usepackage{times}

\shorttitle{Heating time scales in active region cores.}
\shortauthors{Ugarte-Urra \& Warren}

\begin{document}


\title{Determining heating time scales in solar active region cores from AIA/{\it SDO}
  \ion{Fe}{18} images}

\author{Ignacio Ugarte-Urra\altaffilmark{1} and Harry P. Warren\altaffilmark{2}}
\affil{\altaffilmark{1}College of Science, George Mason University. 4400 University Drive,
  Fairfax, VA 22030, USA.}  \affil{\altaffilmark{2}Space Science Division, Code 7681, Naval
  Research Laboratory, Washington, DC 20375, USA.}


\begin{abstract}
  We present a study of the frequency of transient brightenings in the core of solar active
  regions as observed in the \ion{Fe}{18} line component of AIA/{\it SDO} 94\,\AA\ filter
  images. The \ion{Fe}{18} emission is isolated using an empirical correction to remove the
  contribution of ``warm" emission to this channel.  Comparing with simultaneous observations from
  EIS/{\it Hinode}, we find that the variability observed in \ion{Fe}{18} is strongly correlated
  with the emission from lines formed at similar temperatures. We examine the evolution of loops
  in the cores of active regions at various stages of evolution. Using a newly developed event
  detection algorithm we characterize the distribution of event frequency, duration, and magnitude
  in these active regions.  These distributions are similar for regions of similar age and show a
  consistent pattern as the regions age. This suggests that these characteristics are important
  constraints for models of solar active regions. We find that the typical frequency of the intensity
  fluctuations is about 1400\,s for any given line-of-sight, i.e. about 2--3 events per hour. Using 
  the EBTEL 0D hydrodynamic model, however, we show that this only sets a lower limit on the heating 
  frequency along that line-of-sight.
  
\end{abstract}



\section{Introduction}

The rate at which energy is deposited in active region (AR) loops is an important constraint for
heating models of the solar corona.  The case has been made for both low and high frequency
heating, where low frequency implies that the time interval between heating events is longer than
the characteristic cooling time scales of the loops, and it is shorter than the cooling time scale
in the case of high frequency. Low frequency heating has been the standard interpretation
\citep[e.g.][]{cargill1994,cargill2004} of the ``nanoflare model'' in which magnetic field
braiding leads to coronal heating \citep{parker1988}.

\citet{antiochos2003} argued that the absence of cooler loops under the soft X-ray emitting AR
cores implies that there is a source of effectively steady heating that keeps those loops at high
temperatures. The presence of loops at $1\times10^6$ K, evolving (cooling) from higher
temperatures, has, nevertheless, been widely observed in other ARs
\citep{winebarger2005,ugarte-urra2006,ugarte-urra2009,viall2011} suggesting that low frequency
heating is also playing a role. These arguments were later developed in a more quantitative manner
by determining the amount of emission at different temperature ranges and comparing them to model
predictions.  Again, we find in the literature support for both scenarios
\citep[e.g.][]{warren2011,tripathi2011,winebarger2011,warren2012}, suggesting that possibly both
high and low frequency heating can be as important in AR loop formation. In fact, they could
potentially dominate at different instances of an AR long term evolution
\citep{ugarte-urra2012}. Note that some concerns were raised about the limitations of the emission
measure diagnostic \citep{guennou2013} and the implications of these uncertainties in the model
interpretation \citep{bradshaw2012,reep2013}.

In this paper, we address the problem by looking directly to the time evolution of loops in
ARs. We take the simple approach of looking at the intensity fluctuations at the core of active
regions as a measure of the relevant time scales for the heating. We use a newly developed
algorithm to identify relevant events and extract their properties. The discretization of the
lightcurves to extract statistical properties of dynamical features in the corona has been done
multiple times in the past with data from various missions ({\it Yohkoh}, {\it SoHO}, {\it TRACE},
{\it Hinode}), but the focus of most of these studies has generally been in measuring the energies
of the events and estimating their contributions to the overall energy budget in the corona
\citep[e.g.][]{shimizu1995,berghmans1999,nightingale1999,parnell2000,aschwanden2002,li2010}.

We take advantage of the high sensitivity and continuous high cadence observational coverage of
the Extreme Ultraviolet (EUV) solar corona by the AIA instrument on board the {\it SDO} mission.
In particular, AIA provides high cadence spectrally pure images of \ion{Fe}{18}, which has a peak
formation temperature of 7$\times10^6$\,K, and a narrow temperature response, and is well suited to
the characteristic temperatures of AR cores. We also apply a different strategy as we do not group
pixels in space. We look at fluctuations at single pixels to get a characterization of the
short-time dynamics of the overall AR.

Our study shows that the dynamics of the AR can be characterized in terms of several properties
like the frequency of detected intensity enhancements (``events''), their duration and their
amplitudes.  We find distributions that are the same for several mature AR and note that these
distributions change as the AR decays. The characteristic frequency for the events is of about 2--3
events per hour (1400\,s), a time that could be interpreted as the minimum heating frequency along the
line-of-sight.

\begin{deluxetable*}{ccccccc@{$,$}c}
\tabletypesize{\footnotesize}
\tablewidth{0pt}
\tablecaption{AIA/{\it SDO} Observations}
\tablehead{
\multicolumn{1}{c}{Target} &
\multicolumn{1}{c}{Pass} &
\multicolumn{1}{c}{NOAA} &
\multicolumn{1}{c}{Start time} &
\multicolumn{1}{c}{End time}	&
\multicolumn{1}{c}{Fied-of-view}	&
\multicolumn{2}{c}{Center [x,y]} 
}
\startdata
1	&	1    &      11263 &       2011/08/04 20:00 UT	&	2011/08/05 02:00 UT    &       $332\arcsec\times332\arcsec$	&	   $[220\arcsec$ 	&	$167\arcsec]$	\\ 
2	&	1    &      11339 &       2011/11/08 14:00 UT	&       2011/11/08 20:00 UT    &       $332\arcsec\times332\arcsec$	&	   $[0\arcsec$ 		&	$257\arcsec]$	\\ 
	&	2    &      11364 &       2011/12/05 18:00 UT	&       2011/12/06 00:00 UT    &       $332\arcsec\times332\arcsec$	&	   $[-15\arcsec$ 	&	$327\arcsec]$	\\ 
3	&	1    &      11459 &       2012/04/22 00:00 UT 	&       2012/04/22 06:00 UT    &       $302\arcsec\times315\arcsec$	&	   $[160\arcsec$ 	&	$-166\arcsec]$	\\ 
	&	2    &       -  &       2012/05/18 11:00 UT	&       2012/05/18 17:00 UT    &       $280\arcsec\times256\arcsec$	&	   $[64\arcsec$ 	&	$-225\arcsec]$	   	   
\enddata
\label{tab:times}
\end{deluxetable*}

\begin{figure*}[htbp!]
\centering
\includegraphics[width=17cm]{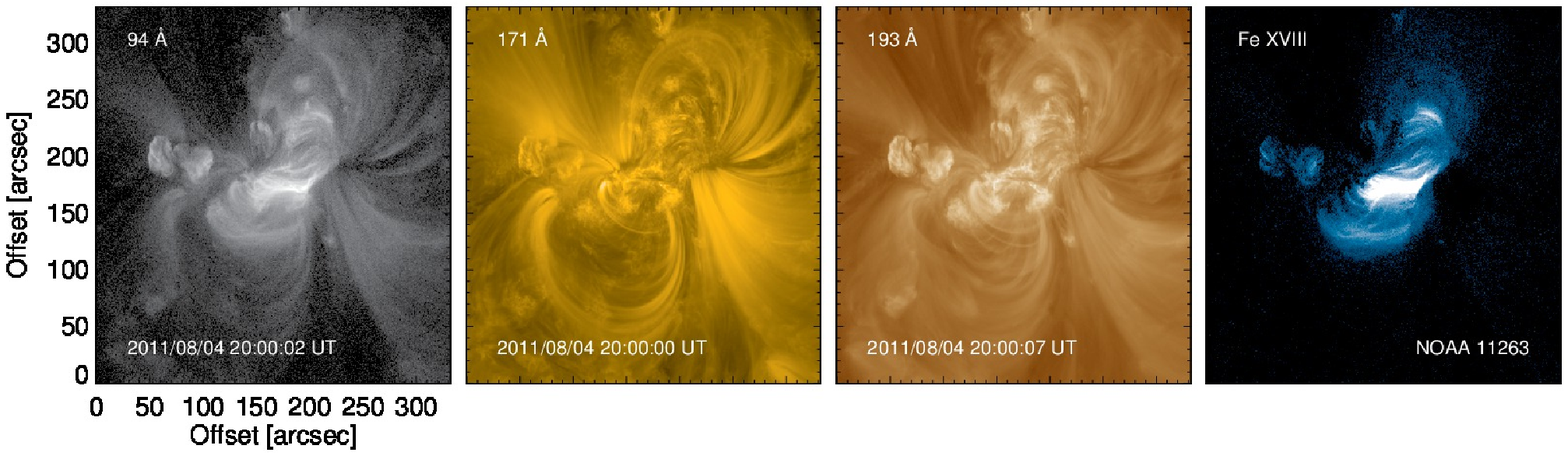}
\includegraphics[width=17cm]{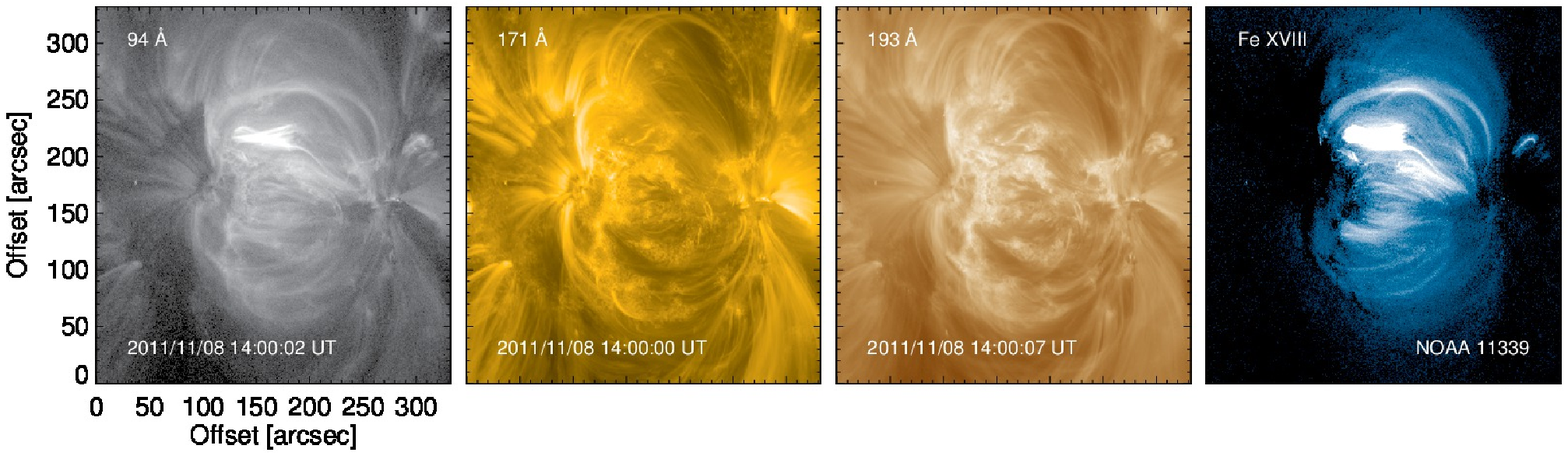}
\includegraphics[width=17cm]{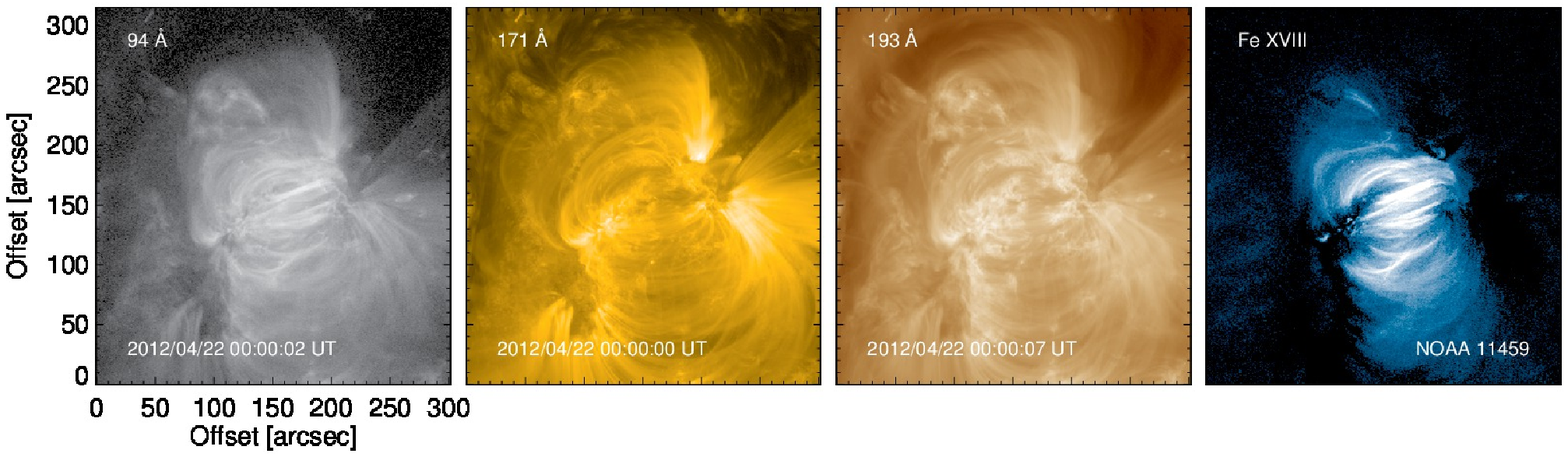}
\caption{Active regions NOAA 11263, 11339 and 11459 as observed by AIA/{\it SDO} in their first
  pass on-disk.  The warm contribution to the 94\,\AA, made out of the 193\,\AA\ and 171\,\AA\
  channels, is subtracted to produce the \ion{Fe}{18} images. }
\label{fig:ARs1}
\end{figure*}
\begin{figure*}[htbp!]
\centering
\includegraphics[width=17cm]{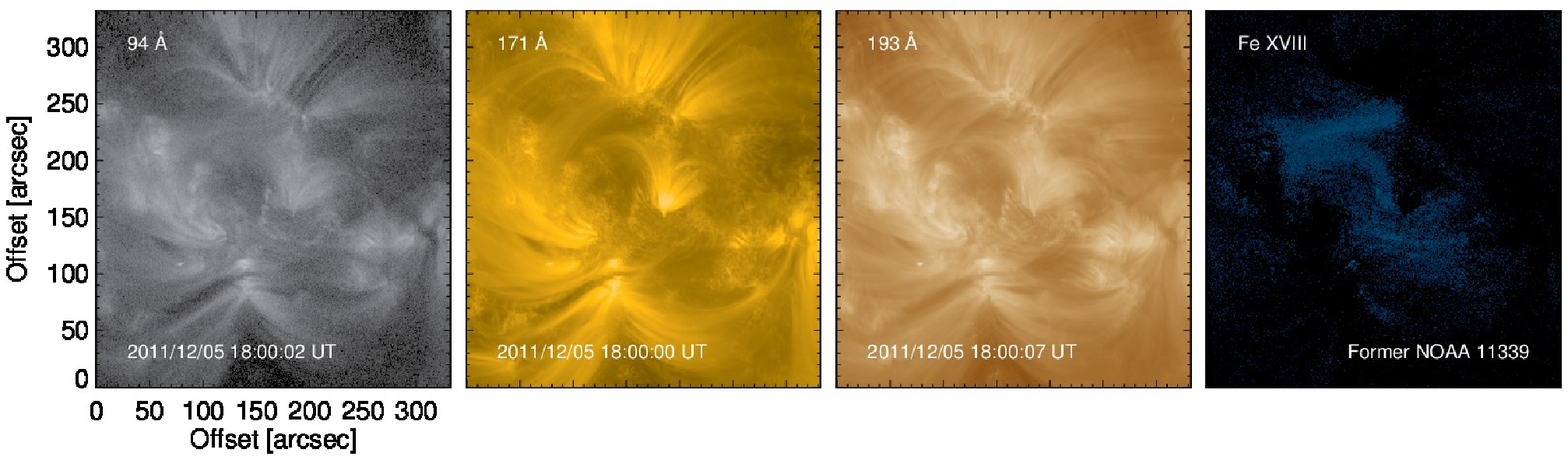}
\includegraphics[width=17cm]{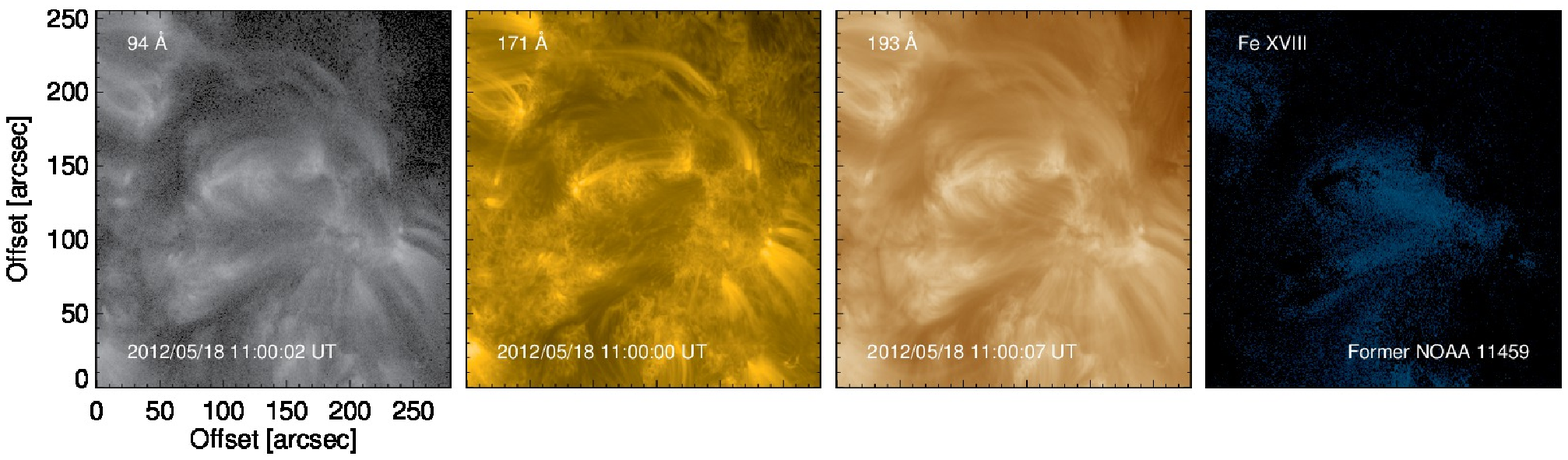}
\caption{Former active regions NOAA 11339 and 11459 as observed in their second pass on-disk, one
  rotation later. NOAA 11459 was labeled 11364 in this pass. Same intensity scaling as in
  Figure~\ref{fig:ARs1}.}
\label{fig:ARs2}
\end{figure*}

\section{AIA/{\it SDO} and EIS/{\it Hinode} observations}
\label{sect:obs}

The Atmospheric Imaging Assembly (AIA) \citep{lemen2012} on board the {\it Solar Dynamics
  Observatory} \citep{pesnell2012} consists of four telescopes that take high-resolution
(0.6\arcsec\ per pixel) images of the full Solar disk in ten narrow-band filters at a cadence of
down to 12\,s.  Seven of those filters image extreme-ultraviolet (EUV) bands covering a
temperature range of 5$\times10^4$\,K to 2$\times10^7$\,K. The 94\,\AA\ band is centered at the
spectral line \ion{Fe}{18} 93.93\,\AA, with a formation temperature of 7$\times10^6$\,K, that can
dominate the emission in active region and flare conditions \citep{odwyer2010}, as it has been
shown in detailed comparisons with spectroscopic observations of active regions
\citep{testa2012,teriaca2012}.

Other ions like \ion{Fe}{10} or \ion{Fe}{14} \citep[CHIANTI,][]{dere1997,landi2013} are also
contributors to the band, becoming even dominant in quiet Sun conditions, which limits the
quantitative diagnostic capabilities of the channel. To circumvent this problem,
\citet{warren2012} devised an empirical method to isolate the ``hot'' \ion{Fe}{18} line emission
from the contaminating ``warm'' component in the channel. The method computes the warm
contribution from a weighted combination of emission from the 193\,\AA\ and the 171\,\AA\
channels, dominated by \ion{Fe}{12} and \ion{Fe}{10} emission respectively.  A method using only
the 171\,\AA\ channel was first considered by \citet{reale2011}.

While these studies demonstrated that the isolated \ion{Fe}{18} emission could be used for a
quantitative analysis of active region images, to our knowledge no investigations have yet
investigated the dynamics of active regions by looking into sets of consecutive images of
\ion{Fe}{18} isolated emission.  From now on we will refer to those images as ``\ion{Fe}{18}
images,'' as opposed to the original ``94\,\AA\ images''.

To demonstrate the applicability of the method to time dependent studies, we compared lightcurves
from sequences of \ion{Fe}{18} images to simultaneous spectroscopic data from the
Extreme-ultraviolet Imaging Spectrometer \citep[EIS,][]{culhane2007} on board {\it Hinode}
\citep{kosugi2007}. The EIS is a high spatial and spectral resolution spectrograph observing in
two wavelength ranges 171--212\,\AA\ and 245--291\,\AA. We investigated sequences of scans of a
60\arcsec$\times$368\arcsec\ area with the 2\arcsec\ slit in 4\arcsec\ steps. Each scan consists
of fifteen 20\,s exposures resulting in a cadence of 300\,s. Data from 15 spectral windows across
the two detectors were retrieved. Here we will present results from a few representative lines
that cover a wide range of temperatures (1.6$\times10^6$\,K -- 4.5$\times10^6$\,K): \ion{Fe}{12} 195.119\,\AA, 
\ion{Fe}{16} 262.119\,\AA, \ion{Ca}{14} 193.874\,\AA\ and
\ion{Ca}{15} 200.972\,\AA. The data were corrected for dark current, cosmic rays, warm and hot
pixels and were calibrated to absolute units using standard software. The intensities of the
spectral lines were obtained from single and multiple Gaussian fits to the lines. Details about
the fits and blends can be found in \citet{ugarte-urra2012}.

\section{Dataset}

We studied the dynamics of AR loops in three active regions: NOAA 11263, 11339 and 11459. The ARs
were chosen from a compiled list of ARs observed by EIS using a particular observing sequence that
includes a large scan with multiple spectral lines for plasma diagnostics and a sequence of fast
scans described in section \ref{sect:obs}. In a previous paper \citet{ugarte-urra2012} showed that
the AR core variability can change as a function of the age of the AR, with emission becoming
fainter and steadier as the AR decays.  Therefore, for the two ARs with a simpler long-term
evolution (11339 and 11459), namely less interaction with neighboring ARs, we also looked at their
dynamics one rotation later. 

Each AR dataset consisted of a compilation of 6 hours worth of 94\,\AA, 193\,\AA\ and 171\,\AA\
level 1.0 images at a 12\,s cadence and limited field-of-view. Table~\ref{tab:times} shows the
dates and times. The images were coaligned using as reference the {\it SDO} master pointing.  To 
improve the signal to noise, images were then averaged to a 60\,s
cadence. The 193\,\AA\ and\,171 \AA\ were then used to compute and subtract the warm component
contribution to the 94 \AA\ channel ($I_{94w}$), resulting in a same cadence sequence of
\ion{Fe}{18} images. This empirical correction is described in \citet{warren2012} and can be
summarized in these expressions
\begin{equation}
I_{94w} = A \sum_{i=0}^3 c_i x^i
\end{equation}
\begin{equation}
x = {fI_{171}+(1-f)I_{193} \over B} < I_{max}
\end{equation}
where $A=0.39$, $B=116.32$, $f=0.31$, $I_{max}=27.5$ and
$c=[-7.19\times10^{-2},9.75\times10^{-1},9.79\times10^{-2},-2.81\times10^{-3}]$.
Figure~\ref{fig:ARs1} shows each AR as seen in the three observed channels, plus the reconstructed
\ion{Fe}{18}.  Figure~\ref{fig:ARs2} shows NOAA 11339 and 11459 in their second pass on disk, one
rotation later, with the same logarithmic intensity scaling.

\begin{figure}[htbp!]
\centering
\includegraphics[width=8cm]{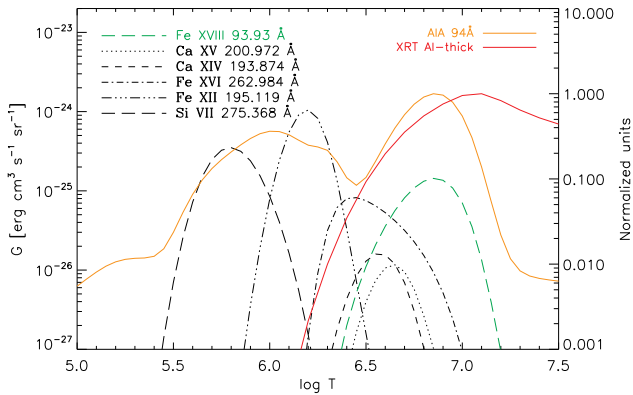}
\caption{Contribution functions for \ion{Fe}{18} 93.93 \AA\ and the spectral lines in the EIS
  dataset together with the normalized temperature response functions of the AIA 94 \AA\ and the
  XRT Al-thick channels.}
\label{fig:response}
\end{figure}

\section{\ion{Fe}{18} time series}

The pre-processing described above opens the door to the analysis of a unique dataset: a 60\,s
cadence time series of isolated \ion{Fe}{18} 93.93\,\AA\ line emission. It is unprecedented at
this resolution and cadence on disk. The SUMER on board {\it SOHO} took spectra of \ion{Fe}{19}
1118 \AA\ at a similar cadence, but in sit-and-stare observations with a narrow slit and off-limb
pointings \citep[e.g.][]{tothova2011}. X-ray imaging has provided imaging at a similar cadence and
resolution for comparable temperatures. The relevance of isolating \ion{Fe}{18} is in its
temperature resolution. Figure~\ref{fig:response} shows the temperature response for the 94\,\AA\
channel, a typical XRT/{\it Hinode} filter and the contribution functions of several spectral
lines including \ion{Fe}{18}.

The figure shows that the XRT filter is the most useful for the study of flare plasma at tens of
million of K.  For AR loops, however, the \ion{Fe}{18} line is, with the Ca lines, the best
diagnostic of dynamics at the temperatures where the emission measure in AR cores peaks, 6.4--6.6
in the logarithmic scale \citep[e.g.][]{warren2012}. Its smaller temperature overlap with cooler
lines, like \ion{Fe}{12} below 6.3, makes it a better discriminator of temperature changes. If we
attempt, as we will in the following sections, to look into the heating time scales of loops by
measuring the time scales between loop brightenings, we need to be able to resolve them in
time. The width of the temperature response is a fundamental factor in the duration of the
brightening. When observing cooling plasma, a broader temperature response implies a longer life
for the brightening, increasing the chances of overlap with preceding and trailing events and our
inability to resolve them.

\begin{figure*}[htbp!]
\centering
\includegraphics[width=8.9cm,angle=90]{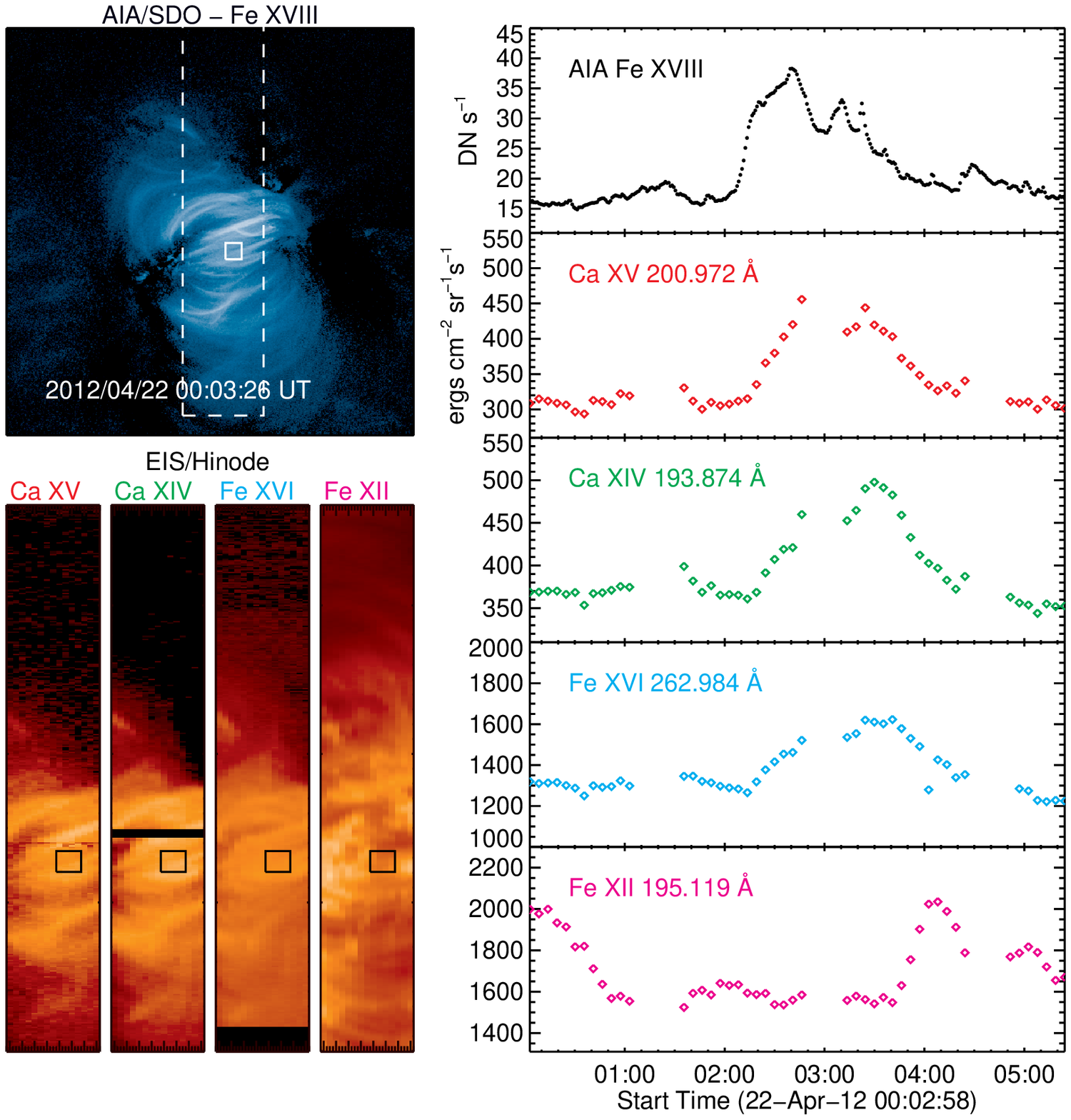}
\includegraphics[width=8.9cm,angle=90]{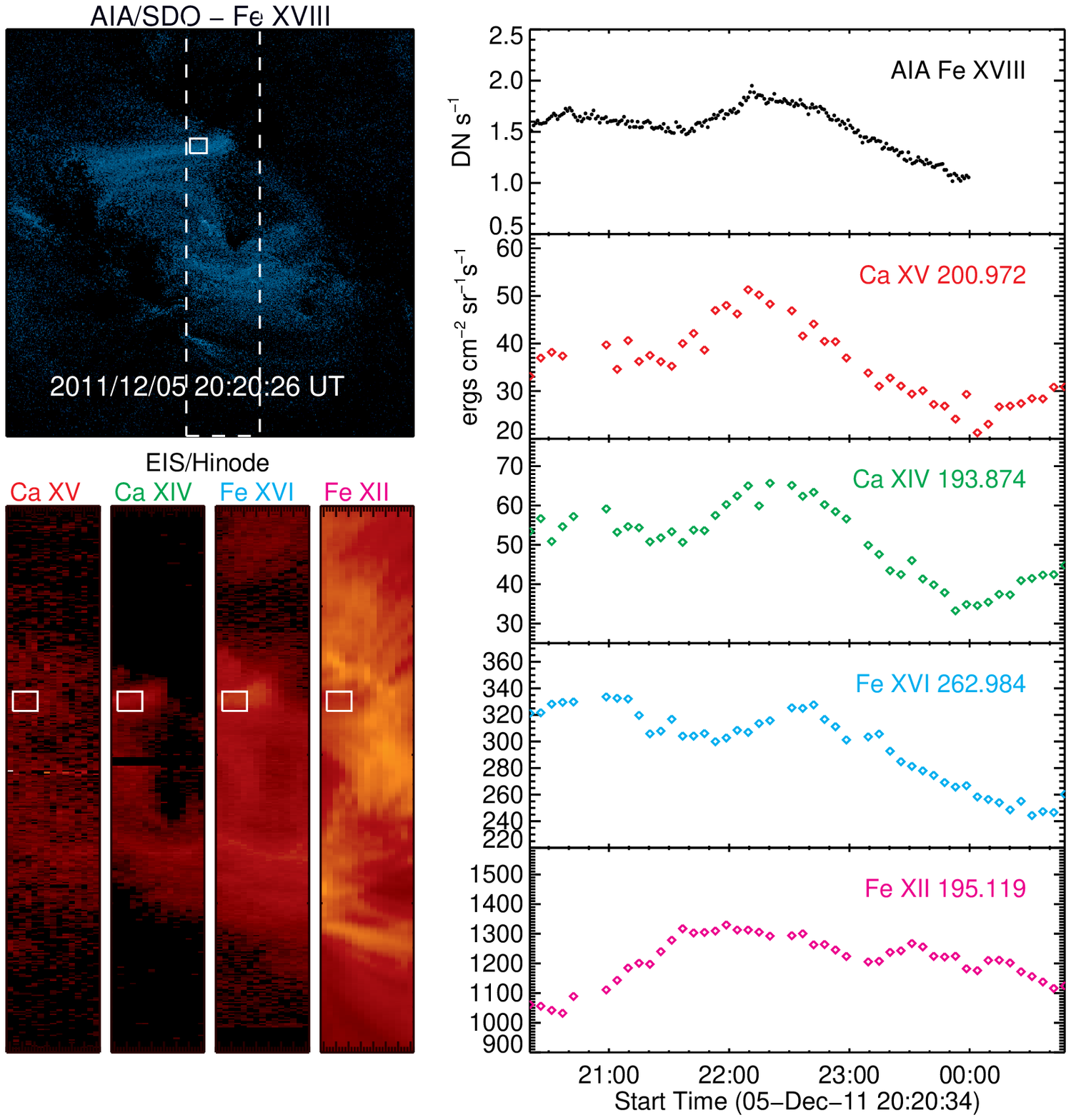}
\caption{Comparison of the AIA/{\it SDO} \ion{Fe}{18} lightcurve to the \ion{Ca}{15--XIV},
  \ion{Fe}{16--XII} ones from EIS/{\it Hinode} for two active regions: NOAA11459 in its first pass on
  disk (left) and NOAA11339 in its second pass (right). A dashed line shows the EIS field-of-view on top 
  of the AIA image. The solid box indicates the area of integration used for the lightcurves.}
\label{fig:lightcurves}
\end{figure*}
\begin{figure*}[htbp!]
\centering
\includegraphics[width=8cm]{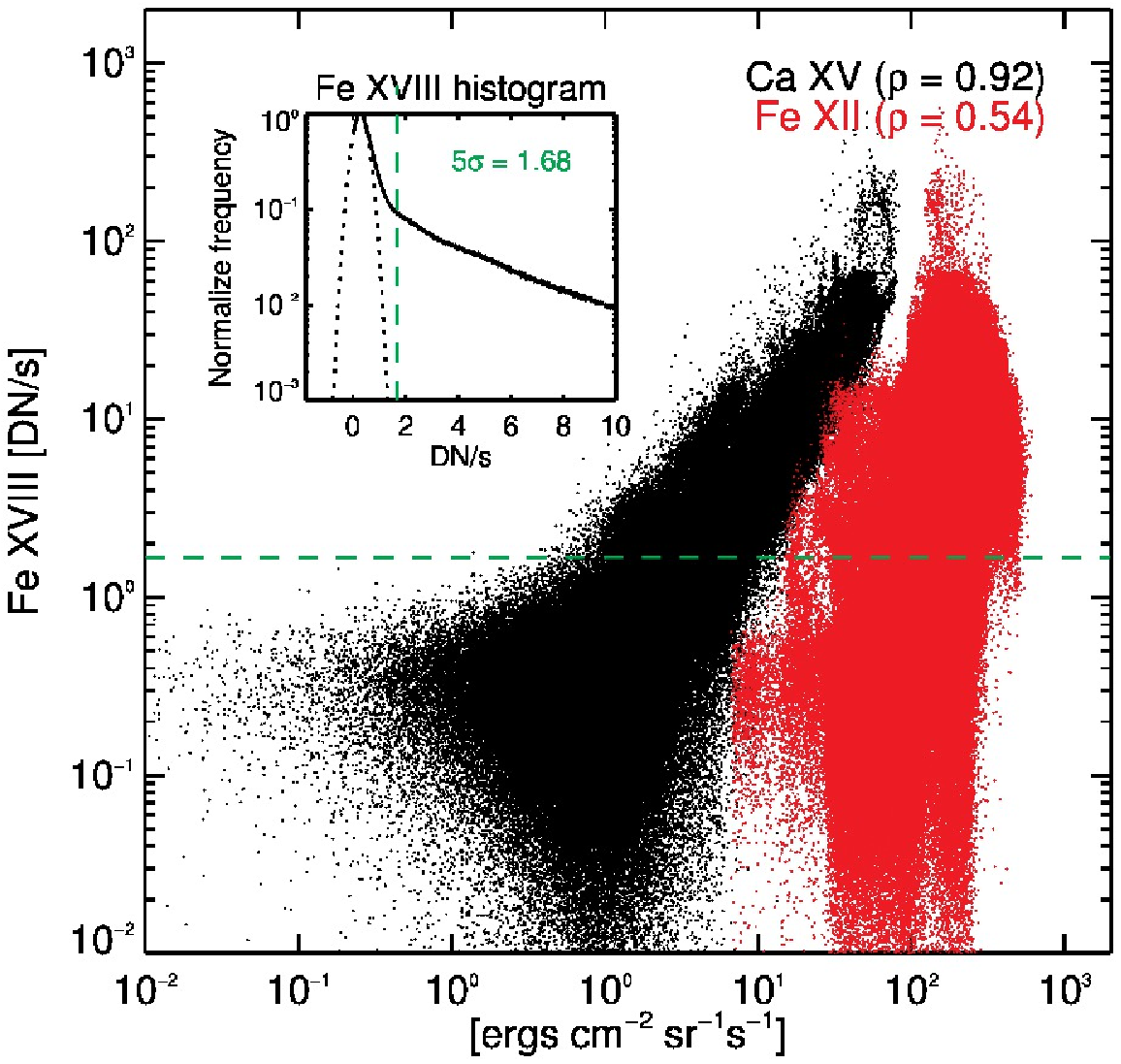}
\includegraphics[width=8cm]{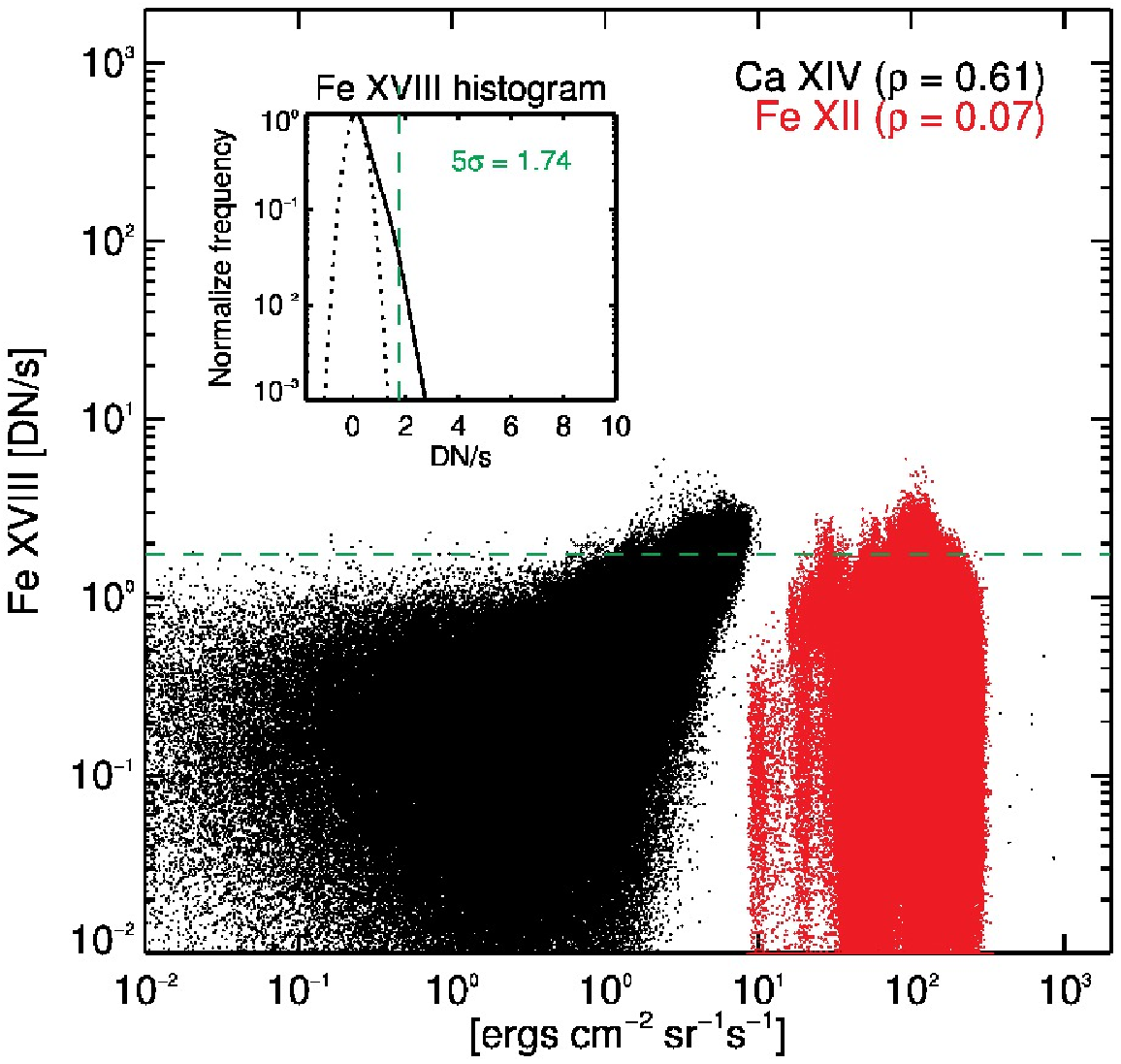}
\caption{Correlation between the EIS intensities and \ion{Fe}{18} for the two datasets shown in
  Figure~\ref{fig:lightcurves}, i.e.  all pixels in the EIS-AIA overlap field-of-view, all
  times. Spearman's correlation coefficient is given in brackets. Inset: histogram for the full
  \ion{Fe}{18} dataset with Gaussian fit (dotted line) of the lower end of the distribution. The
  green dashed line was used as the noise threshold level in the event detection.}
\label{fig:correlations}
\end{figure*}

Similarly, the temperature where the time scale measurements are made is crucial. If the heating
in loops is impulsive, as it is often interpreted from observations \citep[e.g.][]{winebarger2003},
theory predicts that it will heat plasma to multi million degree temperatures in the initial
stages \citep[see reviews by][]{klimchuk2006,reale2010}. As evaporation increases the density in
the corona, the plasma radiates and cools, making the loops become visible in progressively cooler
channels \citep[e.g.][]{winebarger2005,ugarte-urra2009}. If the frequency between heating events
is longer than the time it takes for the loop to cool below e.g. 1$\times10^6$ K, then the
radiative signatures of every heating event are detectable in channels down to those
temperatures. If, however, the frequency is shorter than the cooling time, the heating events will
manifest themselves in a narrower temperature range. While hotter channels, above the cutoff where
the plasma is re-heated, will be able to detect the radiative changes, the cooler ones will remain
blind to that evolution. Therefore, \ion{Fe}{18} observations are more suitable to the study of
heating time scales in AR loops than other cooler lines with a similar narrow temperature
response.

In Figure~\ref{fig:lightcurves} we show sample lightcurves for the 6 hour period of NOAA 11459 in
its first pass on disk and NOAA 11339 in its second. The lightcurves are an integration of a
$15\arcsec\times15\arcsec$ area and are shown in comparison to the lightcurves obtained from the
EIS sequences of the same area. The figure reveals that, similarly to single images
\citep{teriaca2012,warren2012}, the empirical correction does a good job of isolating the hotter
emission in the 94\,\AA\ channel in the time series. The evolution of \ion{Fe}{18} is very similar
to that of \ion{Ca}{14--xv} one and much different than what is observed in \ion{Fe}{12}, meaning
that the warm component of the channel has been successfully
removed. Figure~\ref{fig:correlations} shows the correlation for all pixels in the EIS-AIA overlap
field-of-view and for all times, once the AIA images have been degraded to the EIS resolution. The
correlation is significantly larger when comparing \ion{Fe}{18} and \ion{Ca}{15} than when doing
it with \ion{Fe}{12}. Some correlation is expected in the latter because all the lines will
exhibit larger intensities at the active region than in the quiet Sun. At the 1--2\,DN\,s$^{-1}$
level, the noise starts to dominate (see inset histogram), but the right panel of
Figure~\ref{fig:lightcurves} shows that even variations at that level can be consistent with an
identification of a hot \ion{Fe}{18} component at the core of the active region.

\begin{figure*}[htbp!]
\centering
\includegraphics[angle=90,width=16cm]{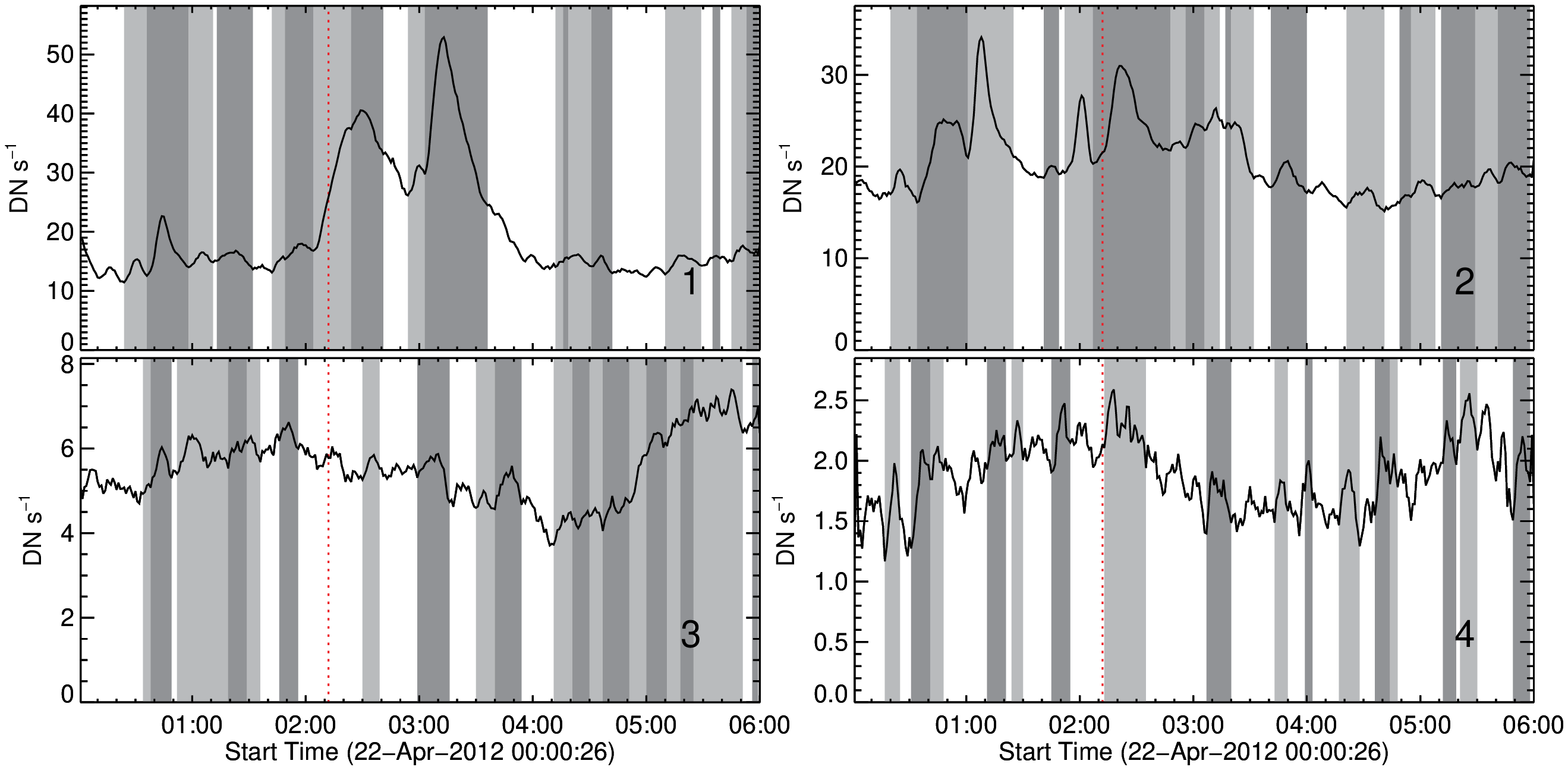}
\caption{\ion{Fe}{18} smoothed lightcurves for the four marked locations in NOAA 11459 in
  Figure\ref{fig:eventsmaps}. Shaded in gray are the events detected by the algorithm. The
  red dotted line indicates the time corresponding to images shown in Figure~\ref{fig:eventsmaps}.}
\label{fig:events}
\end{figure*}
\begin{figure*}[htbp!]
\centering
\includegraphics[width=16cm]{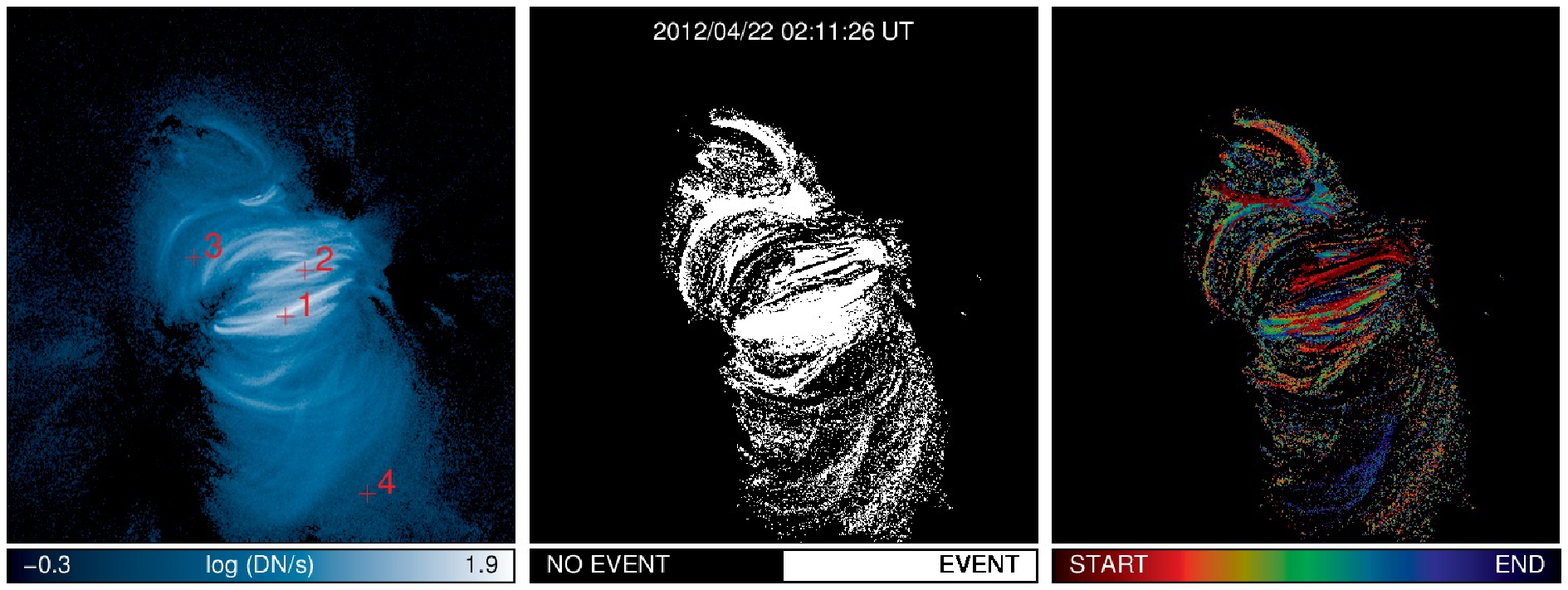}
\caption{Spatial representation of the detection algorithm outputs in the first pass of NOAA
  11459. Left: input intensities; middle: map of detections at a given instant; right: map of the
  events' phases, where time elapsed in event is normalized by duration. }
\label{fig:eventsmaps}
\end{figure*}

\section{Event detection}

Visual inspection of the \ion{Fe}{18} movies reveals a noticeable variability in the loop
structures at the core of the active region, especially when the AR is in its earliest stages of
evolution. There is considerably less variability observed during the second rotation. With the
goal of identifying and characterizing quantitatively the intensity variations of the loops we
devised an algorithm to detect intensity enhancements in single pixel lightcurves.

The algorithm uses threshold levels to establish whether any given set of data points in the
lightcurve conform an event. An event is defined as a sustained (more than three data points)
intensity enhancement over the threshold.  The start, peak and end times of the events are then
saved for future use.

These are the main steps in the detection:
\begin{enumerate}
\item The lightcurves of single pixels for the full spatial domain are smoothed in time with a
  user defined window width ($w$).
\item Each lightcurve is evaluated at every single time step. A time window ($\Delta t$) is used
  to determine the threshold level ($I_{th1}$) at every time step from preceding data
  points. $I_{th1}$ is calculated from the sum of the minimum intensity value within $\Delta t$,
  plus the standard deviation of the intensities within a user defined spatial window around that
  point ($\Delta s$), at the time of the minimum. The evaluation includes determining the trend of
  the intensity (increasing or decreasing) with respect to the last time step.
\item An intensity below the threshold moves the evaluation one time step forward (step 2). An
  intensity above the threshold sets the detection flag on and marks the start of the event and
  determines its peak time (maximum intensity).
\item The peak time is re-evaluated at every time step when the detection flag is on and the trend
  is increasing.
\item In a decreasing trend, the detection flag is set off when the intensity falls below
  $I_{th1}$. Its end time is then recorded. Alternatively, a second threshold $I_{th2}$ is defined
  when a decreasing intensity trend switches to an increasing one resulting in a local minimum. A
  new event is started if the intensity is greater than $I_{th2}$.  New start and peak times are
  then defined. The detection flag is set off when the decreasing intensity falls below $I_{th2}$.
\end{enumerate}
In our analysis $w$ was set to 5 data points (300\,s) to avoid detections due to the one
short-term variability from photon noise. As it will be shown later, this does not fully prevent
the detection of noise. $\Delta t$ was set to 10 data points and $\Delta s$ to 1, so that the
standard deviation is calculated from the eight neighboring spatial pixels. $I(t)+I_{th}$ were
forced to be larger than the noise threshold for the image ($\approx1.8$\,DN\,s$^{-1}$, see
Figure~\ref{fig:correlations}).

Figure~\ref{fig:events} shows representative smoothed lightcurves with detections from the
algorithm for several pixels with different intensity levels in the NOAA 11459 dataset. In those
pixels with high signal to noise (1 and 2) the algorithm does a good job of identifying events
that would have been selected in a visual inspection. As the photon noise becomes more important
(3 and 4), lightcurves become more jagged and the algorithm results only partially agree with a
visual identification. As is discussed in more detail in the next section, some of these detected
events are a result of the noise. This performance is satisfactory for our purposes, as we are
particularly interested in having a characterization of the core of the AR where the \ion{Fe}{18}
signal is high. Whatismore, the ultimate goal is to compare the results of this sort of analysis
done both in observations and full active region simulations.  In such a comparison, the key is
having the same characterization of the dynamics, independently of how successful the
identification of single events is.

Figure~\ref{fig:eventsmaps} shows the results of the detection for a given instant in the first
pass of NOAA 11459. The right panel shows the phase of each event color-coded. The colors are
assigned based on the time elapsed since the start of the event, normalized by the
duration. Elongated stretches of pixels in the same phase demonstrate that the algorithm is able
to detect the evolutionary properties of loop structures from the independent study of single
pixels.

\begin{figure*}[htbp!]
\centering
\includegraphics[angle=90,width=16cm]{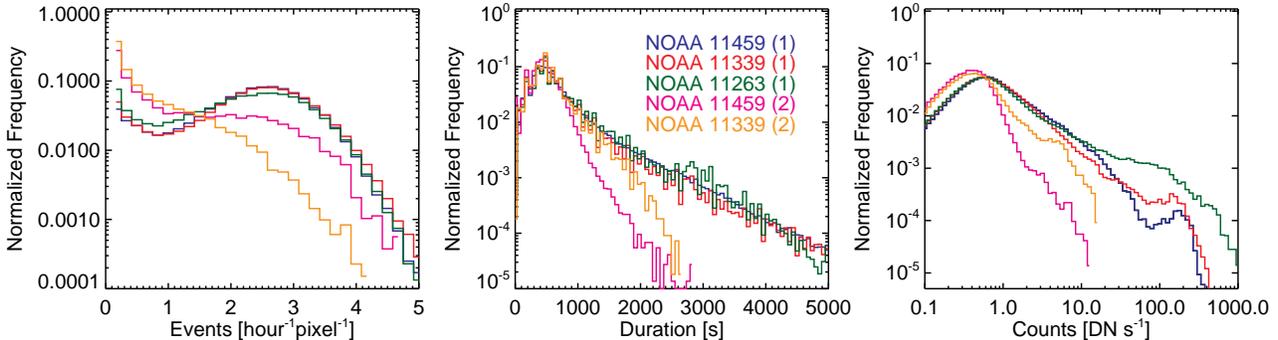}
\caption{Event properties from the analysis of three AR datasets on the first pass on disk and the
  two on the second pass. Left panel: histogram of the number of events per lightcurve at every
  pixel location; middle: distribution of event durations; right: distribution of intensity
  enhancements. Pass number in brackets.}
\label{fig:statistics1}
\end{figure*}
\begin{figure*}[htbp!]
\centering
\includegraphics[angle=90,width=16cm]{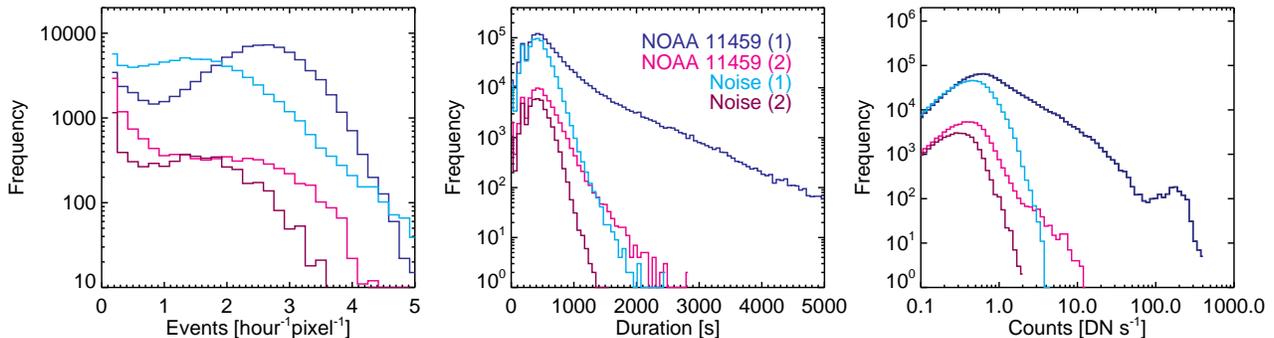}
\caption{Comparison of the properties for events detected on the NOAA 11459 datasets, in two
  stages of evolution, to the properties of events detected from simulated sequences where
  fluctuations are only due to noise.}
\label{fig:statistics2}
\end{figure*}

\section{Events dynamics}

The start, peak and end times of every event, as well as the intensity increase from start to peak
in DN\,s$^{-1}$ are saved for statistical analysis. The number of events, multiple per lightcurve
with several thousand lightcurves per dataset, allows us to look into the statistics of the active
region dynamics as observed in the \ion{Fe}{18} line.

As we are interested in the heating frequency at any location, we first quantified the frequency
of events per lightcurve, namely the number of events per hour at a single pixel
location and for all pixels in an AR.  Left panel in Figure~\ref{fig:statistics1} shows the
histograms for all three AR datasets in their first pass on disk and also in the two with a second
pass. We also determined the duration of all the events and the intensity increase associated to
each of them. These are shown in the middle and right panels of the figure.

The first thing to notice is that all three ARs exhibit very similar properties in the early
stages of evolution, with nearly identical event distributions. Those distributions change
noticeably for the second pass on disk. The typical number of events in the 6 hour time span is
around 15--16 events.  This represents about 2.5 events per hour or an event every 1400\,s. Sample 
lightcurves
1 and 2 in Figure~\ref{fig:events}, at the core, have 18 events each (1200\,s). 
The distributions show that small events are the most
frequent, as well as those that are short. Short and small events can be due to noise fluctuations
in the lightcurves. To investigate the role that the detection of noise could play in our sample,
we constructed a simulated ``noise-only" dataset for each AIA observation, where the base image is
the median intensity image for the time series, and the fluctuations in time are random intensity
additions from a normal distribution centered at 0 and with a standard deviation equal to the
characteristic uncertainty for that intensity. The relationship between different intensity levels
and noise was obtained from the errors, calculated in turn from the standard deviation of the five
12\,s cadence images averaged to make the 1 minute final cadence. We find that estimates of the
uncertainties provided by the standard AIA processing software are conservative with respect to
our own estimate.

Figure~\ref{fig:statistics2} shows a comparison of the results from the detection algorithm in the
AIA data to the results of the simulated data, revealing that, in fact, noise fluctuations are
detected by our algorithm and their characteristics are identifiable in the real data
distributions. They also expose that while noise dominates the fluctuations detected on the low
signal second pass on-disk, there is a distinct variability that can be attributed to intrinsic
variations in the loop structures, with noticeable differences with respect to the early stages of
evolution and reproduceable for different ARs.  Discrepancies in the variability of ARs at the
different stages of evolution had already been noted by \citet{ugarte-urra2012}.

We should, however, be careful with extracting too many conclusions from the duration and count
distributions, because these are statistics for events on single pixel lightcurves. As loops
stretch along multiple pixels, single heating events manifest themselves multiple times in the
distributions given more weight to the properties of intensity enhancements in long loops. Other
studies have addressed this issue by defining events not only by the evolution of the lightcurves
in time, but also in space, grouping pixels with similar evolution. While we find this a valid
approach, we prefer a strategy where there are no assumptions about the relationship between
different volumes of emission. Instead of justifying that every detection of the code is a single
heating event, which could be true in time, but not in space due to the effectiveness of thermal
conduction along the loop, our strategy is to develop a code that can characterize the short-term
dynamics at the core of an AR. This characterization can later be compared to a model of the
atmosphere that attempts to reproduce the observations. Here we have demonstrated that this
characterization can be valid for different active regions attending to the evolutionary stage.

\begin{figure*}[htbp!]
\centering
\includegraphics[width=8cm]{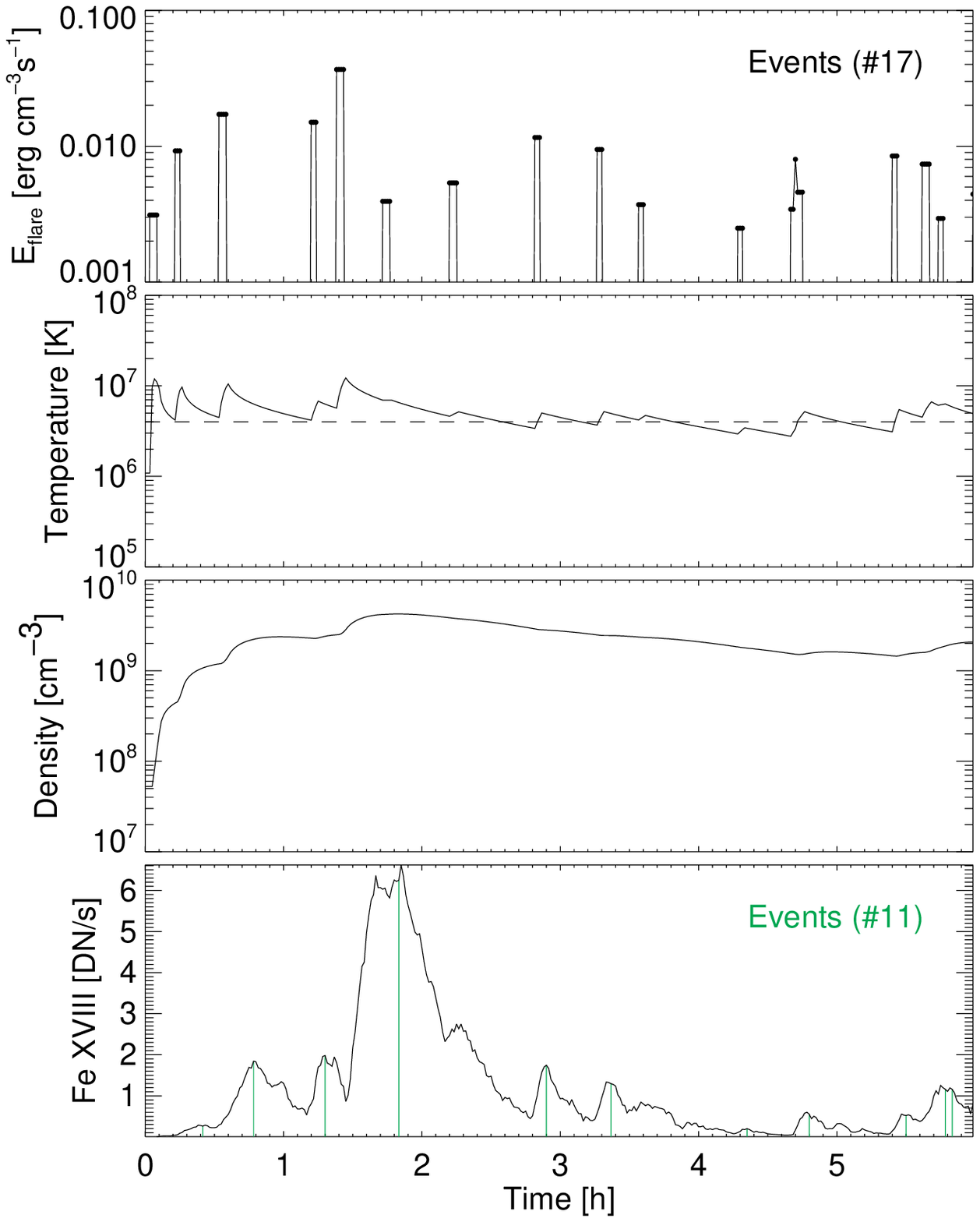}
\includegraphics[width=8cm]{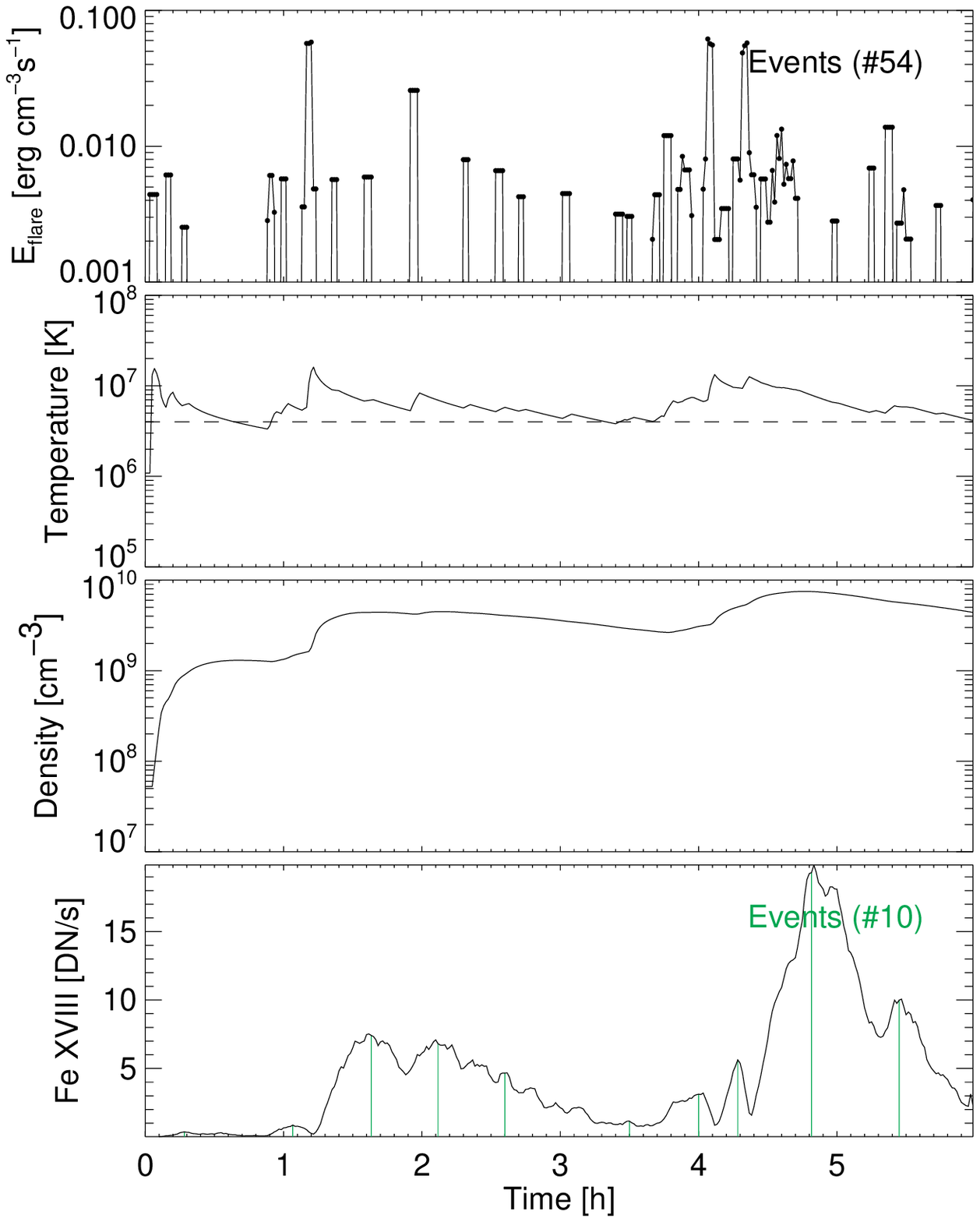}
\caption{EBTEL 0D hydrodynamic simulations of a 100\,Mm loop, 500\,km radius. In the left panel
  case, the heating frequency is comparable to the frequency of events found in the AR
  observations. On the right panel, the frequency is three times larger.  Note the resulting
  \ion{Fe}{18} lightcurves are comparable in terms of smoothness and number of events.}
\label{fig:ebtel1}
\end{figure*}

\section{Discussion}

Our results indicate that there are intensity enhancements at a frequency of around 2--3 events 
per hour along
any given line-of-sight. Interpreting intensity enhancements in loop structures as signatures of
heating events that change the plasma properties in the corona is a common and reasonable
assumption. Observed intensity changes in a particular waveband can be a consequence of density or
temperature variations in the plasma. Both occur in loop evolution. Loops come and go and when
present their electron density is larger than the background corona as measured from spectral line
ratios. Also the presence of temperature variations has been unambiguously determined from looking
at their sequential evolution in different channels sensitive to different temperatures.

To investigate what can be inferred about the heating frequency from these results, we will use a
zero-dimensional hydrodynamic coronal model called ``Enthalpy-based Thermal Evolution of Loops''
\citep[EBTEL;][]{klimchuk2008,cargill2012}. The model, benchmarked with the Hydrad 1D hydrodynamic
code \citep{bradshaw2003}, computes spatially averaged loop properties like the electron density
and temperature, as a function of time, given an ad hoc heating rate function.

As a first test, we computed the temperature and density evolution of a typical loop structure
that is heated repeatedly at a frequency of about 18 events per 6 hour stretch, i.e. 3 events per
hour. We chose a square
heating function with random amplitudes obtained from a power-law distribution $N(E) dE =
E^{\alpha}$, $\alpha=-2$ and a 200\,s event duration. The energies range was set to reach the
typical $4\times10^6$ K temperatures found at the core of ARs. The times were randomly calculated
from a normal distribution of time intervals centered at 1200\,s and a width of 1000\,s. The loop
length was set at 100\,Mm, measured from the footpoints separation of the \ion{Fe}{18} loops and
assuming semicircular shape. To render the \ion{Fe}{18} intensities from the electron density and
temperature, we assumed a loop radius of 500\,km \citep{brooks2012}.

\begin{figure}[htbp!]
\centering
\includegraphics[width=8cm]{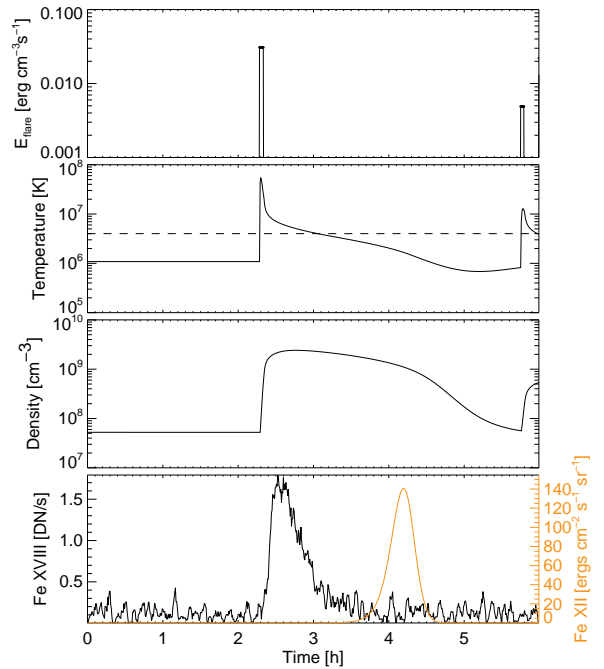}
\caption{EBTEL simulation of a 100 Mm loop with low frequency heating. The bottom panel shows the
\ion{Fe}{18} and \ion{Fe}{12} lightcurves as the loop cools.}
\label{fig:ebtel2}
\end{figure}
\begin{figure}[htbp!]
\centering
\includegraphics[width=8cm]{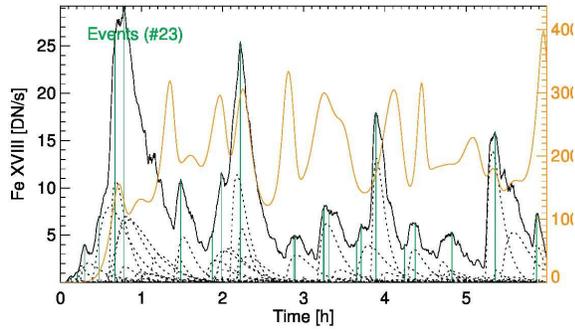}
\caption{Simulated \ion{Fe}{18} and \ion{Fe}{12} envelope lightcurves from the integration of the
  emission produced by 200 low-frequency heating loops along a given line-of-sight. In green the 
  events detected by the algorithm.  The dashed lines represent the \ion{Fe}{18} lightcurves in individual loops.}
\label{fig:ebtel3}
\end{figure}

The left panel on Figure~\ref{fig:ebtel1} shows the time evolution of the density, temperature and
predicted \ion{Fe}{18} counts for a loop heated at that frequency. That cadence of heat deposition
is sufficient to keep the loop at around $4\times10^6$\,K not allowing it cool, making the heating
effectively steady despite its impulsive nature. It is also interesting to note that the number of
events retrieved by the detection algorithm from the \ion{Fe}{18} lightcurve is inferior to the
actual number of heating events needed to produce it. In the simulated observations we see an
envelope of the emission that hides some of the events due to the superposition of the
\ion{Fe}{18} responses to the each of them, plus the smearing due to the noise and the smoothing
window of the processing. This is demonstrated further when we run the same simulation with three
times the number of heating events (right panel in Figure~\ref{fig:ebtel1}) and we obtain a
similar number of detections. We can safely conclude then that the frequency we have measured in
our AR dataset is just a lower limit of the actual heating frequency along the line-of-sight.

It is very unlikely, however, that there is only one loop emitting along a given
line-of-sight. The most likely scenario is that there are multiple loops. We have simulated such a
scenario considering this time loops that are heated impulsively and at a low frequency, a
frequency that allows them to cool down to million degree temperatures (see
Figure~\ref{fig:ebtel2}). We have considered 200 loops heated with the same heating function as
before, but with a frequency of 2--3 random events per six hour stretch. Adding up the
\ion{Fe}{18} emission of all those structures, we obtain an envelope lightcurve
(Figure~\ref{fig:ebtel3}) that resembles the ones described in the single loop test.  The number
of detected events is again smaller, significantly this time, than the actual number of heating
depositions along the line-of-sight.

This comparative analysis of real and simulated lightcurves reinforces the intuitive idea that
counting events from integrated emission along a likely plasma filled line-of-sight, like the core
of an AR, provides only partial clues about the properties of the heating. The line-of-sight
effects and their implications for coronal heating have also been discussed recently by
\citet{viall2013}. Our analysis can be a complement to other diagnostics like the emission measure
distribution (DEM) along the line-of-sight.  The DEM has already been used as a potential
discriminator for the heating frequency by establishing the different contribution levels for the
emission and different temperatures very dependent on whether the plasma is heated repeatedly and
not allowed to cool or more sporadically allowing for emission at all temperatures \citep[see
discussions in][]{warren2011,tripathi2011,mulu-moore2011,ugarte-urra2012,bradshaw2012}.

\section{Conclusions}

We present the analysis of unique spectrally pure \ion{Fe}{18} AIA/{\it SDO} high spatial
resolution and high cadence observations of three ARs in different stages of evolution. The
formation temperature of this line makes it an ideal candidate to study the properties of the
heating time scales at the core of AR.  We demonstrate our ability to isolate the line and study
the intensity fluctuations in time down to very low count rates. We then construct an algorithm to
quantify the observed variability by counting the number of transient brightenings (events), that
could be a proxy for heating events.

We find that the ARs investigated show reproduceable traits in their variability like the number
of detected events during the same span of time, resulting in a characteristic frequency of about
2--3 events per hour (1400\,s) for any given line-of-sight view.  We show that, while this is a
constrain on the maximum time between heating events, the frequency of the heating could be
significantly higher, as we can only detect events from the envelope lightcurve. This envelope can
be the result of the integration of many smaller events.

Furthermore, these detected events have duration and strength distributions that are very similar
for different ARs, making them a potentially testable constrain for full active region time
dependent models. We, therefore, suggest that a promising avenue to circumvent the inherent
limitations of characterizing the time scales of the heating through line-of-sight integrated
emission difficult to deconvolve, is to apply the statistical analysis techniques used on real
observations to the forward modeling results of full active region heating.


\acknowledgments
The authors acknowledge funding from NASA grant NNX13AE06G through the ROSES 2012 Program NNH12ZDA001N-SHP.
AIA data is courtesy of NASA/SDO and the AIA science team. 
Hinode is a Japanese mission developed and launched by ISAS/JAXA, with NAOJ as domestic partner and NASA 
and STFC (UK) as international partners. It is operated by these agencies in co-operation with ESA and NSC 
(Norway). 
CHIANTI is a collaborative project involving George Mason University, the University of Michigan (USA) 
and the University 
of Cambridge (UK).


\bibliography{thispaper.bib}
\bibliographystyle{apj}

\end{document}